\documentclass[11pt,a4paper]{article}
\usepackage[T1]{fontenc}
\usepackage[utf8]{inputenc}
\usepackage{lmodern,microtype}
\usepackage[margin=25mm,headheight=15pt]{geometry}
\usepackage{graphicx,booktabs,tabularx,array,amsmath,amssymb}
\usepackage[font=small,labelfont=bf]{caption}
\usepackage{enumitem,fancyhdr,xcolor,url}
\usepackage[section]{placeins}
\usepackage[colorlinks=true,linkcolor=black,citecolor=black,urlcolor=blue!45!black]{hyperref}
\hypersetup{pdftitle={TeamLens in Critsly: A Consent-Based Team-Composition Interface and Synthetic Readiness Evaluation for Design Collaboration},pdfauthor={Nizam Kadir},pdfsubject={Technical preprint; not peer reviewed}}
\graphicspath{{figures/}}
\setlist{nosep,leftmargin=*}
\newcolumntype{Y}{>{\raggedright\arraybackslash}X}
\newcommand{\release}{\texttt{2448c7f}}
\newcommand{\fullrelease}{\texttt{2448c7fdaf4cdebfcae47d10ef0579234ea222f9}}
\newcommand{\artifact}[1]{\texttt{\detokenize{#1}}}
\newcommand{\studyfigure}[5]{%
  \begin{figure}[!htbp]\centering
  \includegraphics[width=#3\linewidth,height=#4\textheight,keepaspectratio]{#1}
  \caption{#5}\label{#2}
  \end{figure}}

\title{\textbf{TeamLens in Critsly:}\\[0.3em]
A Consent-Based Team-Composition Interface\\
and Synthetic Readiness Evaluation\\
for Design Collaboration}
\author{Nizam Kadir\\[0.4em]
\normalsize Science, Mathematics and Technology\\
\normalsize Singapore University of Technology and Design, Singapore\\
\small \href{mailto:nizam_kadir@mymail.sutd.edu.sg}{nizam\_kadir@mymail.sutd.edu.sg}\\
\small ORCID: \href{https://orcid.org/0000-0002-6725-1133}{0000-0002-6725-1133}}
\date{\small Technical preprint; not peer reviewed\\25 September 2026}

\begin{document}
\maketitle
\thispagestyle{plain}

\begin{abstract}
\noindent
Discussing working preferences may support reflection within a design team, but a personality label should not become a performance prediction or a condition of participation. This technical report presents TeamLens, an optional Critsly interface for voluntarily sharing a self-reported MBTI type with a particular board. It separates account activation from disclosure, displays descriptive composition counts, and provides distinct controls for disabling visibility, withdrawing one report and deleting all of one's reports. The evaluation combines released-source inspection, independently specified synthetic aggregation cases, access and lifecycle checks, browser component tests, a local database microbenchmark and deployment records. All 65,536 binary eligibility subsets of sixteen fixed, distinct type reports matched an independent oracle; 128 seeded multiplicity fixtures also matched, after 4,259 synthetic share calls. The isolated HTTP suite passed 183 assertions. In 180 sequential in-memory SQLite read trials, median service-call time increased from 0.064\,ms with no profile rows to 200.932\,ms with 1,024 rows; instrumented SQL operations followed $11+3n$. These findings concern exercised software behaviour and a bounded workload. They do not establish human usability, psychometric validity, learning gains, team-performance effects or production capacity. The contribution is an implemented disclosure-to-aggregation workflow and an auditable technical account of its correctness boundaries, privacy limitations and scaling cost. OpenAI Codex assisted with implementation, evaluation and manuscript preparation; the paper discloses this use and its limits.
\end{abstract}


\section{Introduction}

Design teams bring together people with different ways of discussing ideas, examining evidence and organising work. A digital workspace can make those differences discussable without treating a categorical personality label as an explanation of a person's competence. The design problem addressed here is therefore modest: allow collaborators to describe a preference if they choose, see a bounded summary of shared reports, and retain control over their disclosure.

TeamLens extends Critsly's board workspace with that workflow. Each account starts with the feature disabled. Enabling it reveals a board-side panel but does not create a personality report. A named board owner or editor may then submit a type they already know, with an explicit statement that the type will contribute to that board's counts. The panel describes the resulting composition and suggests inclusive discussion practices. It neither administers an MBTI assessment nor infers a type from drawings, dialogue or other board activity.

\subsection{Platform context and motivating study}

Critsly has previously been presented as a board-grounded critique environment combining a visual canvas, guided reflection, perspective-based AI critique and reusable critique traces~\cite{critsly}. That system presentation provides platform context; it does not evaluate TeamLens or establish learning gains. TeamLens shares Critsly's board infrastructure with StudioCrit, but its evaluated behaviour concerns voluntary disclosure, exact aggregation, eligibility and withdrawal. It does not reuse StudioCrit's critique-classification distributions, export schema or simulated account counts as TeamLens measurements. The present evaluation reports fresh observations from the released TeamLens implementation.

Hendra and colleagues studied 535 students in 108 self-formed design teams and reported a weak association between distinct MBTI types and project grades ($r=0.189$, $p=0.0457$)~\cite{hendra}. The team is the relevant unit for that association. Their single-course observational study did not directly assess teamwork quality or establish causation. Its questionnaire-derived types and participants are external evidence, not a TeamLens dataset. Here it motivates cautious reflection rather than a predicted grade, preferred team composition or replication claim.

This report contributes an implemented account-to-board disclosure workflow; a separation of retained reports, eligible aggregates and deletion; fresh deterministic and seeded synthetic tests; and a bounded examination of read-path cost. All newly generated observations are software observations. No students were recruited, no questionnaires or interviews were administered, and no human collaboration or learning outcome was measured.

\section{Research Aim and Design Foundations}

Four engineering questions organise the evaluation:
\begin{description}[font=\normalfont\bfseries,leftmargin=3.2em,labelwidth=2.7em]
\item[RQ1.] Does TeamLens return exact board-scoped counts for the defined valid-type and eligibility fixtures?
\item[RQ2.] Do activation, explicit sharing, updates, disabling, re-enabling and withdrawal produce the specified stored and visible states?
\item[RQ3.] Do the exercised authentication, tenant, CSRF, membership and own-record boundaries prevent unintended operations or disclosure of other members' identity--type pairings?
\item[RQ4.] What interface, release and read-path evidence supports further controlled use, and what remains unmeasured?
\end{description}

The questions concern implementation and observability. Successful execution cannot establish that showing type counts helps a team, that a type is stable or accurate, or that particular types should work together. McCrae and Costa's examination of MBTI indices found no support for qualitatively distinct types or genuinely dichotomous preferences in their sample~\cite{mccrae}. This older study does not evaluate TeamLens, but reinforces the decision to treat four-letter entries as reported categories rather than validated natural classes or measures of ability.

The Myers \& Briggs Foundation's ethical guidance emphasises voluntary participation, control over disclosure and nonjudgmental interpretation~\cite{ethics}. It is normative guidance from the instrument organisation, not evidence of educational effectiveness. TeamLens records self-reports; it does not claim certification, official assessment administration, endorsement or independently validated compliance with that guidance.

Groupware research treats awareness support as an interface-design problem. Gutwin and Greenberg distinguish knowledge about collaborators, mechanisms for maintaining that knowledge and its uses during shared work~\cite{gutwin}. Their framework concerns ongoing interaction in a shared workspace. TeamLens instead displays voluntarily reported composition: it is an adjacent awareness interface, not a measure of ongoing activity or coordination. This connection motivates an inspectable display without establishing that personality categories improve collaboration.

Bellotti and Sellen frame privacy in collaborative systems through feedback and control over information capture, subsequent processing, accessibility and purpose~\cite{bellotti}. Applied to TeamLens, this perspective distinguishes account activation, board-specific disclosure and withdrawal. The sharing action and aggregate result are visible, while exclusion and deletion have separate controls. This is an interpretation of an established framework, not evidence that users understand the interface or that every privacy criterion has been satisfied.

Three design commitments follow. First, opting into the interface and opting into a board's counts are separate actions. Second, the displayed denominator is the set of eligible shared reports, not the roster or every person involved in a project. Third, users can remove their own reports even when normal composition viewing is unavailable. Aggregation reduces direct disclosure of identities in the response, but does not guarantee anonymity: in a small team, prior knowledge and a one-person change can reveal a contributor's type. This residual risk is stated in the sharing interface. The artefact contribution is the implemented and tested separation of activation, disclosure, current eligibility, retained storage and withdrawal in a collaborative board. It applies established awareness and privacy concerns to an explicit stored-state and visible-state contract; it does not propose a newly discovered privacy principle.

\section{System Design and Implementation}

\subsection{Account preference, board scope and interface}

The evaluated implementation is pinned to \release; the full commit is recorded in the \hyperref[sec:availability]{data and software availability statement}. TeamLens runs in Critsly's shared PHP application runtime and uses existing board-access rules. It is a Critsly feature, independently controlled from StudioCrit. The service identifies the tenant, authenticates the actor, checks the account preference and resolves named board access before returning composition. Figure~\ref{fig:architecture} shows the implemented boundary.

\studyfigure{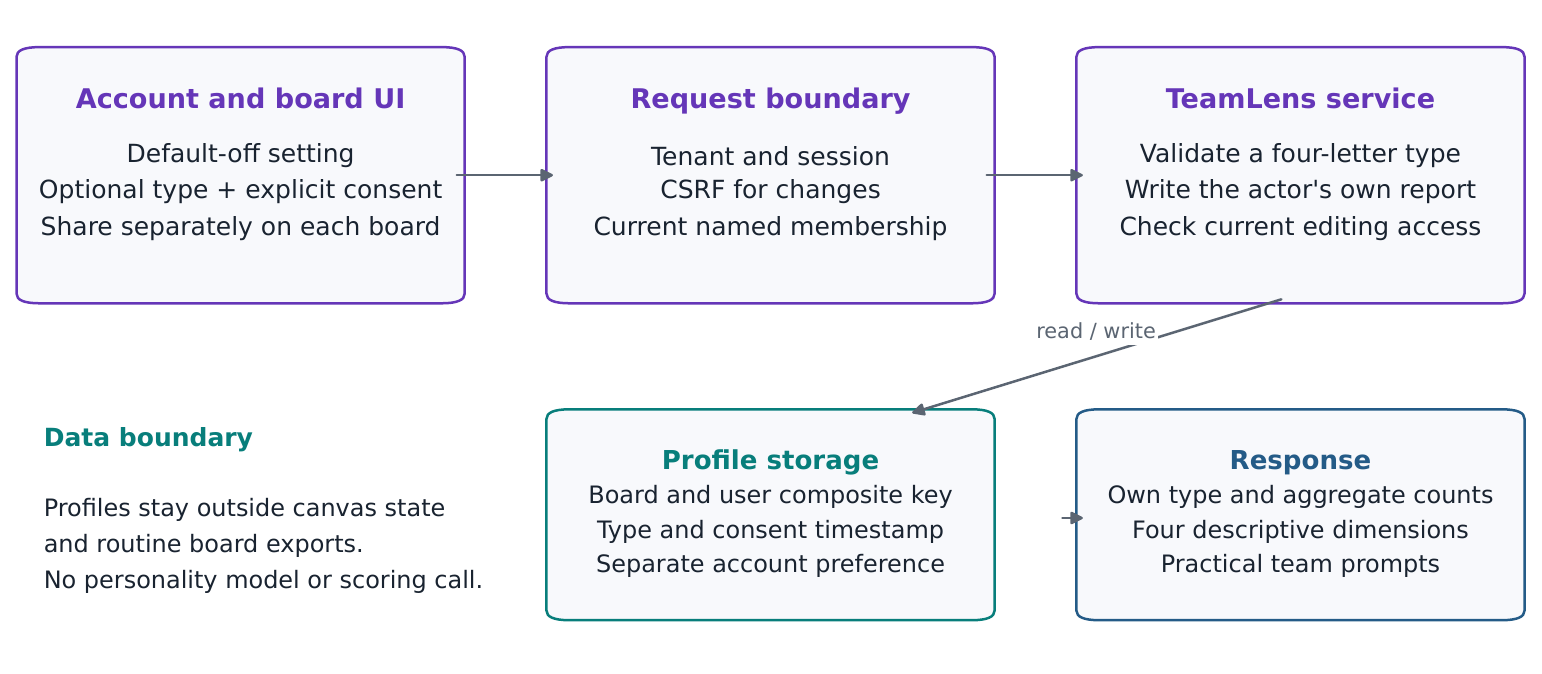}{fig:architecture}{1.00}{0.31}{TeamLens architecture and data boundary. Account activation, board-scoped self-reports, access checks and aggregate rendering are distinct stages. The feature does not infer personality from canvas content or send these reports through an AI classification pipeline.}

In Settings, users enable TeamLens under Board Preferences and save the preference. Existing boards must reload to receive that setting. The board sidebar then offers TeamLens, with sections for the user's working preferences, team composition, practical discussion prompts and the research source (Figures~\ref{fig:settings}--\ref{fig:sharing}). A named owner or editor can share; a named view-only collaborator can inspect counts. The existing critique-only board view hides the sidebar, so API permission and visible UI access should not be conflated.

\begin{table}[!htbp]
\caption{Implemented scope and its interpretive boundary.}
\label{tab:scope}\small
\begin{tabularx}{\linewidth}{@{}p{0.20\linewidth}YY@{}}
\toprule
Layer & Implemented capability & Boundary \\
\midrule
Account & Default-off setting; explicit save; preserved by older clients that omit the field. & Activation does not create a report. \\
Board disclosure & Self-entered valid type; explicit consent; update or withdrawal. & Only the authenticated actor's report is written. \\
Composition & Eligible report count, distinct types, type histogram and four paired dimensions. & No score, rank, target ratio or inference about nonresponders. \\
Access & Critsly tenant, session, named membership, edit permission for sharing and CSRF for writes. & Public-link canvas access is insufficient. \\
Data control & Disable/re-enable, one-board withdrawal and deletion of all own reports. & Hiding or exclusion is distinct from erasure. \\
\bottomrule
\end{tabularx}
\end{table}

\studyfigure{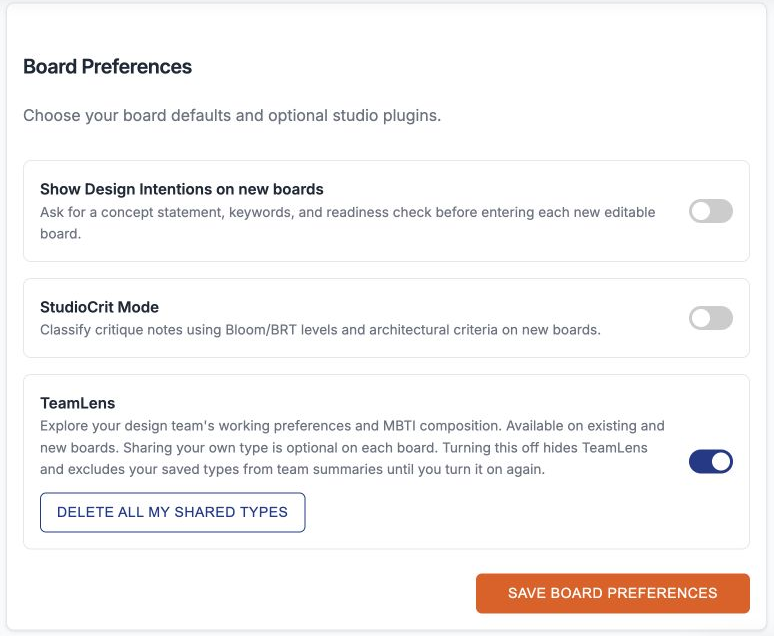}{fig:settings}{1.00}{0.43}{TeamLens in Critsly Board Preferences after the final settings-layout correction. This is an authentic production screenshot of a demonstration account. The switch and Save Board Preferences control govern account activation; sharing remains a separate board action.}

\studyfigure{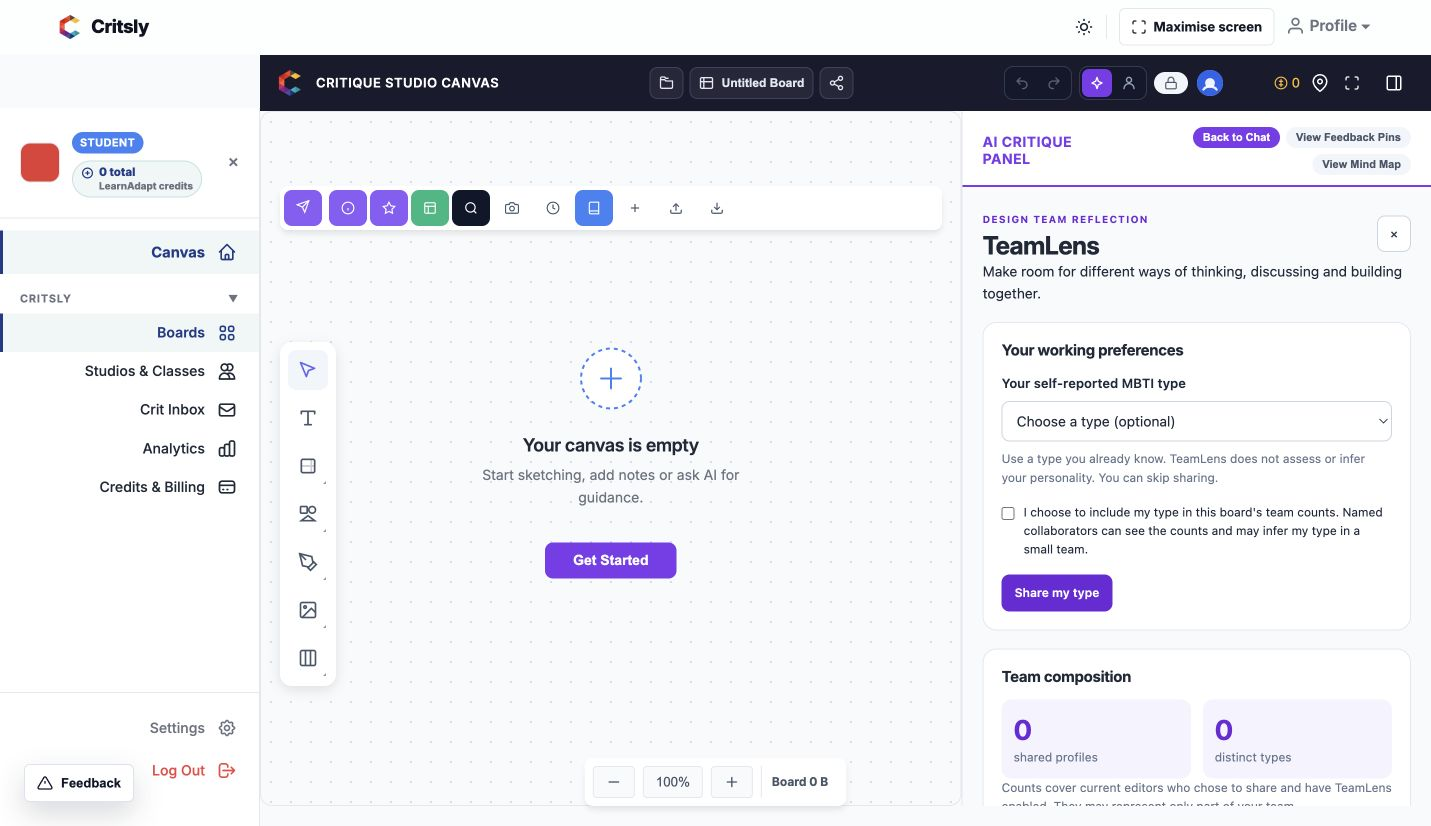}{fig:board}{0.98}{0.30}{Production board with the TeamLens panel open. The empty board and demonstration account are interface evidence, not a participant session or design-learning observation. The panel can be inspected without submitting a type.}

The screenshots use actual production interfaces and synthetic demonstration data. Any displayed INTJ entry was entered for illustration, not inferred about the user or recommended as a desirable type. The sharing form explains small-team inference before submission. A successful save identifies the actor's own reported type; changing it requires another explicit submission. Counts refresh when the panel opens, not through continuous polling, so an already open view may remain stale until reopened.

\studyfigure{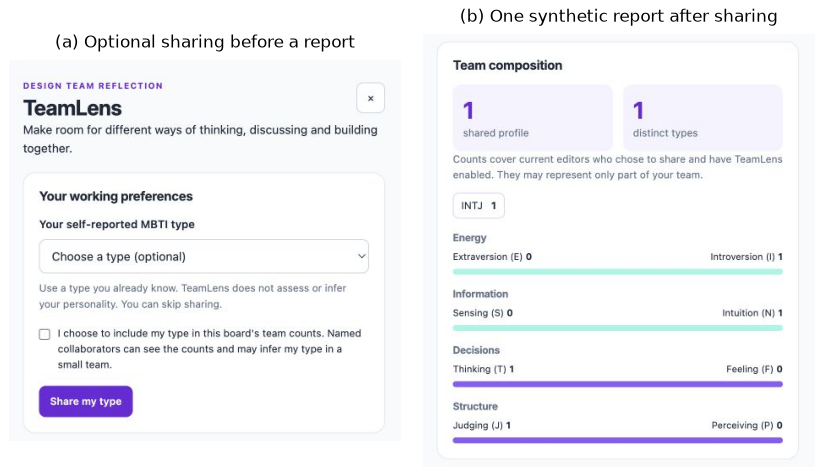}{fig:sharing}{1.00}{0.42}{Optional sharing and resulting composition in the production interface: (a) the unsubmitted type/consent form and (b) the team-composition panel containing one shared profile. The displayed INTJ value is synthetic demonstration data, manually entered rather than inferred. The counts are an interface example, not a participant-study result.}

\FloatBarrier
\subsection{Storage, eligibility and response fields}

A dedicated table stores one report for each board--user pair. The composite primary key makes a later share an update rather than a duplicate contribution. The user's setting is stored separately. Table~\ref{tab:fields} lists the source-level fields relevant to interpretation; it is not a dump of production records.

\begin{table}[!htbp]
\caption{Stored state and returned fields in the released implementation.}
\label{tab:fields}\small
\begin{tabularx}{\linewidth}{@{}>{\raggedright\arraybackslash}p{0.23\linewidth}>{\raggedright\arraybackslash}p{0.32\linewidth}Y@{}}
\toprule
Location & Fields & Meaning \\
\midrule
Account preference & \artifact{teamlens_mode_enabled} & Account activation flag; named access and editing eligibility are checked separately. \\
Private report table & \artifact{board_id}, \artifact{user_id}, \artifact{type_code}, \artifact{consented_at} & Board-scoped self-report, actor identity and submission timestamp; not a research-participation consent record. \\
Actor response & \artifact{ownType}, \artifact{canShare} & Current actor's report and permission to share. \\
Aggregate response & \artifact{participantCount}, \artifact{uniqueTypeCount}, \artifact{typeCounts}, \artifact{dimensions} & Eligible-report denominator, distinct categories and exact marginal counts. No other-user profile list is returned. \\
Withdrawal response & \artifact{success}, \artifact{ownType} & Completion without returning board composition. \\
\bottomrule
\end{tabularx}
\end{table}

For every read, stored reports are joined to current users and filtered by the enabled preference. Each candidate's current named board role is then checked; only owners and editors with a valid stored type code contribute. Invalid stored codes are skipped. Public-link visibility does not supply named membership. Current account email is used in membership resolution rather than trusting an email supplied by the caller or retained in an older session. The response contains the requesting user's own type and aggregate counts, without attaching names or user identifiers to the histogram.

If $E_b$ is the set of eligible stored reports on board $b$, the displayed profile count is $N_b=|E_b|$. For a type $t$, $c_b(t)$ counts its occurrences in $E_b$, and the distinct-type count is the number of $t$ with $c_b(t)>0$. Each paired dimension sums the corresponding letters across those same reports. Thus the type histogram and every dimension pair must sum to $N_b$. The API field \artifact{participantCount} denotes reports in this implementation; it must not be mistaken for recruited research participants.

\subsection{Disabling, withdrawal and deletion}

Disabling TeamLens retains saved reports but excludes them from summaries. Re-enabling can restore a report's contribution if the account retains eligible access. Per-board withdrawal deletes only the actor's row for that board; deleting all shared types removes that actor's rows across boards. These deletion operations require authentication, the Critsly tenant and CSRF protection, but deliberately do not require an enabled setting or continued board membership. They return no composition to an actor who may no longer be entitled to view it.

Reports are kept separately from canvas state rather than embedded in shared nodes. This separation does not itself prove that every future integration will preserve the boundary, and private storage still contains identifiable account references. The present report evaluates exercised response and lifecycle behaviour, not a complete information-flow or retention audit.

\section{Evaluation Method and Evidence Provenance}

\subsection{Frozen software and distinct evidence layers}

Fresh local tests used the clean checkout at \release, with source hashes and command manifests retained. The environment included PHP 8.5.4, Node.js 24.4.1 and Chrome 153.0.8010.53. The HTTP harness copied application source into an operating-system temporary directory, used a fresh SQLite database, and ran actual PHP front-controller and authentication routes over loopback. It excluded existing databases, environment files and inherited cloud/database credentials. Service tests and the larger synthetic experiments used separate in-memory databases.

The browser component harness exercised the actual panel JavaScript with simulated responses; its 13 mock requests are not 13 production transactions or independent users. Separately retained staging/production smoke summaries each record 26 checks, while the final release audit links source provenance, container digest, ready revisions and asset hashes. These layers have different evidential roles and are not pooled into a single sample (Table~\ref{tab:evidence}).

\begin{table}[htbp]
\centering\small
\caption{Evidence layers and analytical units. Counts are kept separate; none represents recruited participants.}
\label{tab:evidence}
\begin{tabularx}{\linewidth}{@{}>{\raggedright\arraybackslash}p{.25\linewidth}>{\raggedright\arraybackslash}p{.32\linewidth}X@{}}
\toprule
Layer & Unit and extent & Recorded result \\
\midrule
Service smoke & One isolated service suite & Passed; case count not inferred \\
HTTP integration & 183 assertions over actual PHP routes & All passed in isolated SQLite fixture \\
Browser component & One scripted workflow, 13 simulated API requests & Passed; desktop and narrow views \\
Eligibility enumeration & 65,536 subsets of 16 stored types & Zero complete-summary mismatches \\
Multiplicity fixtures & 128 seeded fixtures; 4,259 share calls & Zero final-summary mismatches \\
Access and lifecycle & 40 explicit service cases; 8 synthetic users, 2 boards & Zero expected-state mismatches \\
Read microbenchmark & 180 timed calls; 30 excluded warmups & Six sizes, 30 timed repetitions each \\
Staging live smoke & 26 checks on staging.critsly.com & Passed; generated board and reports removed \\
Production live smoke & 26 checks on critsly.com & Passed; generated board and reports removed \\
\bottomrule
\end{tabularx}
\end{table}

\subsection{Complete binary eligibility domain}

The first aggregation experiment fixed sixteen synthetic editing members, one for each valid four-letter type, and a separate enabled owner with no report. The sixteen reports were created through the service's share path. A 16-bit mask then toggled only the members' enabled preferences. All $2^{16}=65{,}536$ masks were evaluated. Expected counts were specified independently using fixed feature bitmasks and binary population counts; complete-summary equality was checked, together with the observer's absent own type and sharing permission.

This is exhaustive coverage of a precisely defined binary eligibility domain over sixteen fixed, distinct reports. It is not exhaustive coverage of possible inputs, sequences, roles, data corruption, concurrent requests or deployment states. The count and distinct count are equal by construction in this experiment, which is why a second experiment is needed to exercise repeated categories.

\subsection{Seeded multiplicity fixtures and boundary tests}

A separate Python generator, seeded with 20260925, produced 128 frozen fixtures and expected summaries. The first fixtures were empty, one of each type and two of each type. Remaining fixtures selected a size from zero through 64, sampled a type pool and drew repeated types from it. Expected histograms used Python's Counter and an explicit feature mapping; no TeamLens aggregation code was imported. Each fixture occupied its own synthetic board, with enabled editing accounts and a separate non-reporting owner. Reports were created through actual share calls before the final summary comparison.

These fixtures test multiplicity and the distinction between report count and distinct-type count. They are constructed test cases, not a probability sample of natural teams. The fixed seed and retained type lists support reruns; the generated type frequencies should not be interpreted as plausible population prevalence.

Access and lifecycle tests complemented those aggregation experiments. A dedicated direct-service runner evaluated forty explicit cases across eight synthetic users and two boards, including owner, explicit permission, invitation and studio-membership paths. It makes no HTTP or CSRF claim; those boundaries were exercised by the separate actual-route harness. Together the tests covered default-off behaviour; explicit boolean consent; valid and invalid types; missing or wrong CSRF; anonymous and other-tenant requests; view-only and unlisted accounts; public links; own-record writes despite a supplied foreign user identifier; membership revocation; and withdrawal while disabled or after losing access. Settings saves checked both explicit toggles and preservation when the preference was absent or submitted through another tenant. Expected status and stored/visible state were checked rather than treating any HTTP response as success.

\subsection{Sequential read microbenchmark}

Read-path cost was measured at $n\in\{0,4,16,64,256,1024\}$ eligible stored profiles. A fresh in-memory SQLite database was created for each size, processed in ascending order. A separate enabled owner read the board without contributing a profile. Member profiles and edit permissions were inserted before timing, representing already stored synthetic reports; the benchmark does not measure consent submission. Types cycled through a fixed Cartesian ordering and were equally represented at sizes divisible by sixteen.

Each size received five warm-up reads and thirty timed reads: 180 measured service calls and thirty excluded warm-ups. Timing used the monotonic \artifact{hrtime(true)} clock around \artifact{TeamLensService::read}. Result validation and serialisation were outside the interval; PDO operation counters were inside it, with detailed SQL tracing disabled. Median time is the mean of the fifteenth and sixteenth sorted trials; p95 is the nearest-rank twenty-ninth observation out of thirty.

The runtime was PHP 8.5.4 with CLI OPcache disabled, SQLite 3.51.3 and ARM64 macOS/Darwin 25.5.0, on an Apple M2 Max with twelve CPU cores and 32\,GiB of memory. The machine was not isolated from other interactive or operating-system work. The fixture mirrored query-relevant source columns and membership indexes; it did not inspect production index availability or reproduce MySQL storage, collation or network cost. Calls were sequential in one process, without HTTP or Cloud Run. Consequently, these measurements describe a local instrumented microbenchmark, not production latency, concurrent throughput or classroom capacity.

\section{Findings}

\subsection{Aggregation correctness within the defined domains}

Every one of the 65,536 eligibility masks matched the complete independent expected summary, with zero observed mismatches. The run took 24.586\,s in its local harness. Across the 128 multiplicity fixtures, 4,259 actual synthetic share calls preceded the comparisons, and all final summaries matched. That run took 4.382\,s. These elapsed values describe entire correctness harness phases, not the latency measure in Section~\ref{sec:performance}.

\begin{table}[htbp]
\centering\small
\caption{Observed summaries for three selected eligibility masks. All rows retain 16 synthetic stored reports and match the independent oracle exactly.}
\label{tab:experiments}
\begin{tabularx}{\linewidth}{@{}Xrrrrrr@{}}
\toprule
Eligibility & Profiles & Types & E/I & S/N & T/F & J/P \\
\midrule
All enabled & 16 & 16 & 8/8 & 8/8 & 8/8 & 8/8 \\
E types enabled & 8 & 8 & 8/0 & 4/4 & 4/4 & 4/4 \\
None enabled & 0 & 0 & 0/0 & 0/0 & 0/0 & 0/0 \\
\bottomrule
\end{tabularx}
\par\vspace{3pt}{\footnotesize These constructed distributions illustrate participation filtering. They are not observed student distributions or a recommended team composition.}
\end{table}

Figure~\ref{fig:participation} presents selected observed masks from the complete-domain run. Sixteen stored reports can yield zero, a subset or sixteen eligible contributions without deleting any row. This directly illustrates why stored-report count, displayed denominator and team roster are different quantities. A zero count or missing category does not establish nonresponders' types or the whole team's composition. The absence of mismatches supports the implementation's arithmetic for these fixed domains; it does not validate what a self-reported personality label means.

\studyfigure{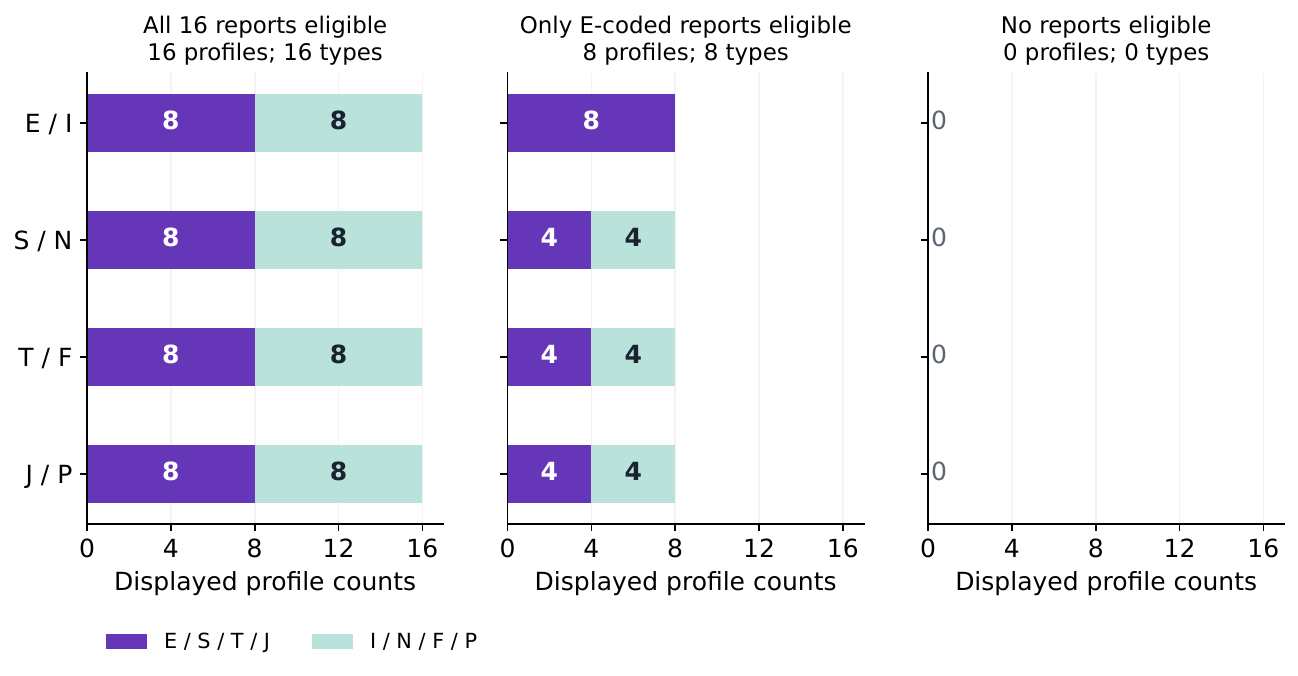}{fig:participation}{0.95}{0.26}{Observed selected eligibility masks from the retained synthetic results. The underlying sixteen stored type reports are fixed while account-level eligibility changes. Counts are deterministic software outputs for synthetic fixtures, not the distribution of types in a recruited population.}

\subsection{Workflow, access and lifecycle observations}

The service smoke suite completed successfully, and all forty dedicated service access/lifecycle cases matched their expected states. The actual-route HTTP suite passed 183 assertions, including session authentication, tenant separation, CSRF rejection, board membership, strict consent and type validation, preference persistence and preservation, and actor-bound deletion. The count is the harness's assertion count: multiple assertions can concern the same request or state transition. It is not a sample size or an estimate of the fraction of all possible defects excluded.

The HTTP fixtures confirmed that disabling removed a report from another eligible member's aggregate without erasing it, re-enabling restored it, and membership revocation excluded an old report. A revoked or disabled member could still withdraw their own record. A caller-supplied other user identifier did not redirect sharing or deletion. Account-wide withdrawal removed the actor's reports across boards while preserving another account's report. The response inspection found no other-user identity fields in the exercised composition payloads, and the tested canvas states remained unchanged.

\studyfigure{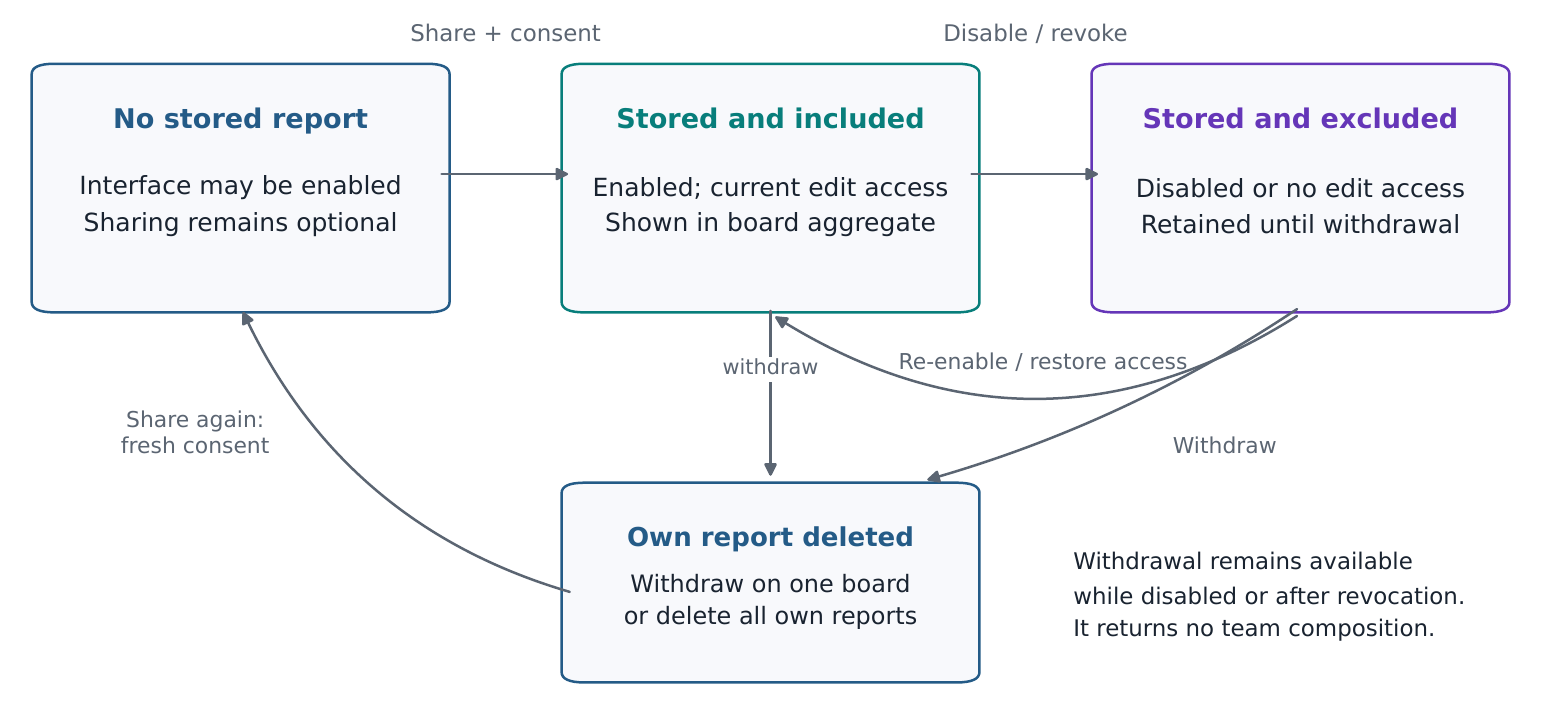}{fig:access-lifecycle}{1.00}{0.34}{Released access and data-lifecycle model, checked against the exercised cases. Preference state, eligibility and report deletion are separate. Withdrawal does not grant composition access and operates on the authenticated actor's reports. The diagram is a code-grounded model, not an exhaustive security proof.}

The component browser harness passed while issuing 13 simulated requests and recording desktop/narrow viewport evidence. Production screenshots establish that the relevant controls were rendered in the live interface; they do not measure task difficulty, comprehension or satisfaction. In particular, no human usability score is reported. Functional observations are strongest for the specific states and layouts exercised, and a response-level privacy check does not eliminate small-team inference or every possible data path.

\subsection{Read-path timing and SQL growth}\label{sec:performance}

Table~\ref{tab:performance} reports the measured series. Median read time increased from 0.064\,ms at $n=0$ to 0.758\,ms at sixteen profiles, 20.224\,ms at 256 profiles and 200.932\,ms at 1,024 profiles. At the largest size, p95 was 210.417\,ms and the maximum was 219.893\,ms. These rounded values are derived from retained per-trial measurements, not estimated from the source code.

\begin{table}[htbp]
\centering\small
\caption{Local sequential read microbenchmark. Each row summarizes 30 timed calls after five warmups on a fresh in-memory SQLite fixture. SQL operations include prepared-statement execution, queries and schema checks.}
\label{tab:performance}
\begin{tabular}{rrrrr}
\toprule
Profiles $n$ & Median (ms) & p95 (ms) & Maximum (ms) & SQL operations \\
\midrule
0 & 0.064 & 0.070 & 0.078 & 11 \\
4 & 0.245 & 0.322 & 0.504 & 23 \\
16 & 0.758 & 0.813 & 0.914 & 59 \\
64 & 3.229 & 3.533 & 3.598 & 203 \\
256 & 20.224 & 20.923 & 21.376 & 779 \\
1,024 & 200.932 & 210.417 & 219.893 & 3,083 \\
\bottomrule
\end{tabular}
\end{table}

The instrumented number of SQL operations was exactly $11+3n$ at all six tested sizes. Here an operation means a prepared-statement execution, direct query or exec call, including schema inspection and \texttt{CREATE TABLE IF NOT EXISTS}; preparing and executing one statement are not counted twice. The three per-profile operations correspond to repeated membership resolution. There were eleven operations at zero profiles and 3,083 at 1,024 (Figure~\ref{fig:performance}).

\studyfigure{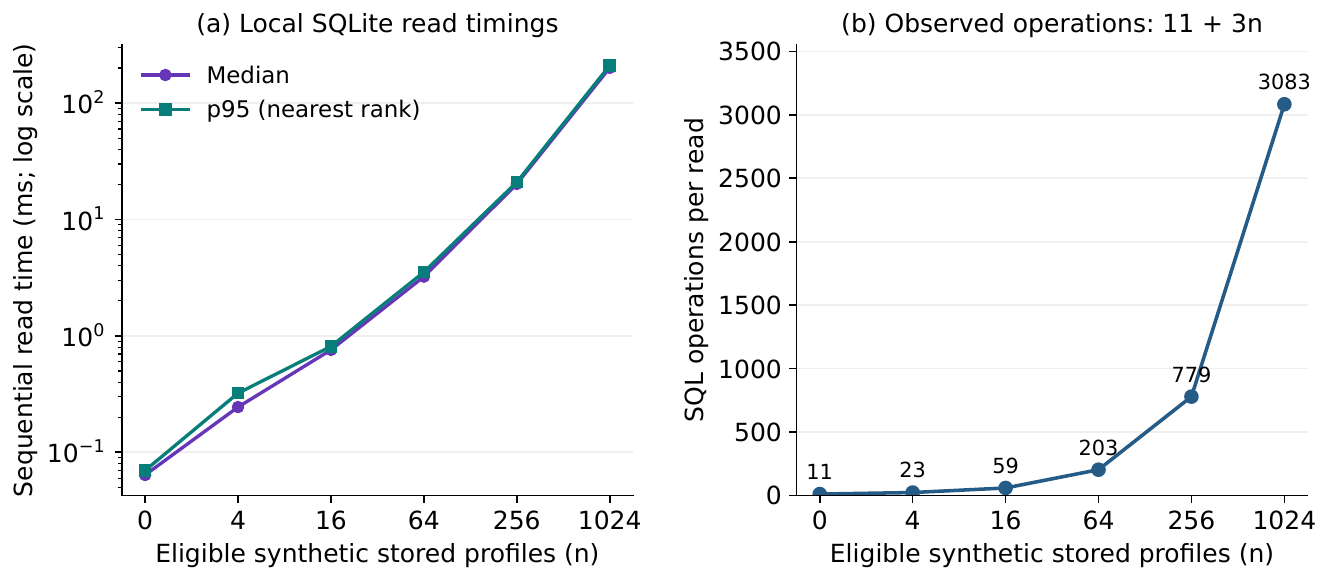}{fig:performance}{1.00}{0.29}{Actual local sequential read timings and instrumented SQL operations across six synthetic profile counts. Each size has thirty measured trials after five warm-ups. SQL growth is $11+3n$ in this fixture; the timing curve is not a Cloud Run latency or capacity estimate.}

The observed time increase is steeper than the linear operation count over the larger fixtures. Operation count alone therefore does not determine runtime: query work, lookup plans, row counts and instrumentation all matter. Current per-profile access rechecking serves a useful purpose, because revoked membership must stop contributing, but incurs repeated database work. The results identify a candidate for future optimisation and regression testing; they do not demonstrate a production bottleneck or the benefit of an unimplemented optimisation.

\subsection{Deployment evidence and production interface}

The final audit records a successful root-context build of \release and matching source provenance: no uploaded untracked files or committed-content mismatches were recorded. The resulting container image digest was used by both Critsly services in project \artifact{learnadapt-platform}, region \artifact{asia-southeast1}. Staging and production remained separate deployments with their tenant configuration preserved, apart from release metadata. Table~\ref{tab:deployment} identifies their audited revisions.

\begin{table}[htbp]
\centering\small
\caption{Final isolated Critsly deployments verified on 25 September 2026. Both run the same immutable image and Git revision.}
\label{tab:deployment}
\begin{tabularx}{\linewidth}{@{}Xlll@{}}
\toprule
Environment & Revision suffix & Live checks & Traffic \\
\midrule
Critsly staging & \texttt{00209-b5m} & 26/26 passed & 100\% \\
Critsly production & \texttt{00153-98d} & 26/26 passed & 100\% \\
\bottomrule
\end{tabularx}
\par\vspace{3pt}{\footnotesize Commit: \texttt{2448c7fdaf4cdebfcae47d10ef0579234ea222f9}. Image digest: \texttt{e67ecc993147...33eaff6204}. Full identifiers and timestamps are retained in the release audit; other product services were not deployed.}
\end{table}

\studyfigure{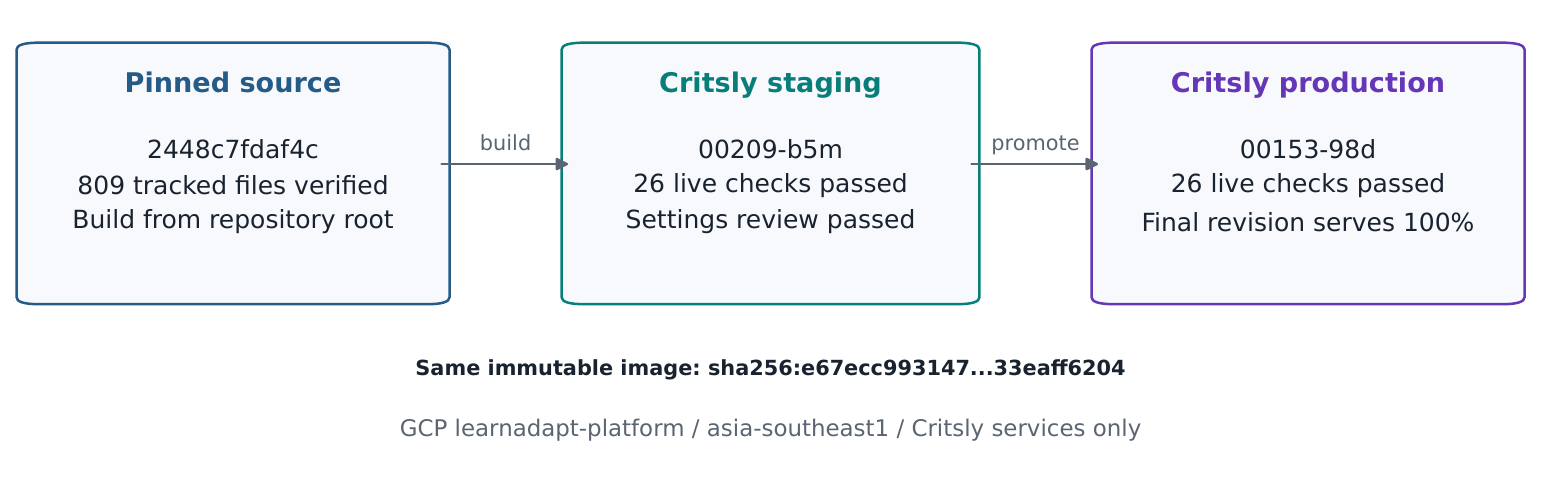}{fig:deployment}{0.95}{0.23}{Recorded Critsly release path from the pinned root-context source to the shared image digest, separate staging verification and production revision. This is a release-provenance record for the named services, not a deployment claim for other LearnAdapt applications.}

The final audit linked deployed-SHA markers and JavaScript/CSS asset hashes to the release and observed authentication rejection for anonymous TeamLens access. Post-promotion functional smoke summaries report 26 successful checks at each Critsly environment and removal of the synthetic test board and shared types. The run manifest records completion at approximately 07:15:21 UTC on 25 September and links each output hash to its target revision. The smoke output itself does not embed a Git SHA; the separate build/revision audit and run manifest establish that linkage rather than a live URL alone.

Production interface captures are dated and labelled as demonstrations. Some board-panel captures precede the final settings-only CSS correction; the relevant board JavaScript and CSS hashes are unchanged. Settings captures show the corrected final layout. This provenance matters because an authentic screenshot demonstrates a particular interface state, not every behaviour of the running service or a complete user journey.

\section{Discussion and Limitations}

\subsection{An inspectable, voluntary composition workflow}

The strongest supported contribution is a working separation between enabling a feature, sharing one report, inspecting an eligible aggregate and withdrawing data. Independent-oracle agreement, actual-route checks and interface evidence make that separation inspectable. They provide a firmer engineering basis than screenshots alone, while preserving the distinction between a descriptive tool and a validated educational intervention.

Optional participation also limits interpretation. Those who share may differ from those who do not, and disabling, access changes or withdrawal alter the denominator. A board with four eligible reports cannot automatically be described as a four-person team. The interface correctly identifies shared profiles, but users may still overinterpret labels or infer a colleague's type from context. Clear wording and deletion controls address parts of that risk; this study does not measure whether people understand or respect those boundaries.

\subsection{Validity and generalisation}

The exhaustive claim is deliberately narrow. Binary eligibility over sixteen fixed types covers every mask in that construction, but not arbitrary type multiplicities, all access combinations, input encodings, concurrent updates or failures. Seeded multiplicity fixtures broaden arithmetic coverage without turning the evaluation into a population study. The oracles are independently implemented but can still contain conceptual mistakes; retaining their explicit mappings enables inspection and further verification.

Neither a valid four-letter string nor exact aggregation establishes the reliability of a self-report. No assessment was administered, and the instrument provenance or stability of a real person's label was not evaluated. There is no evidence here that TeamLens improves design quality, inclusion, learning or team performance. The discussion prompts remain practical suggestions rather than interventions validated by either these tests or the cited design-team study.

The local SQLite microbenchmark excludes network, authentication middleware, rendering, cold starts, remote database latency and concurrent users. Its ascending order, small trial count, warm-up policy and shared workstation further constrain interpretation. Deployed MySQL indexes were not inspected. SQL-operation growth is a directly observed implementation property in the fixture; transferring its measured times to a production service would be unjustified.

\subsection{Privacy, governance and further evaluation}

Aggregate output should not be called anonymous when group size, known membership or before/after observations allow inference. The stored board--user relationship is identifiable, even if the response omits a named list. This evaluation did not audit every backup, log, retention setting or integration. It reports implemented deletion and the tested state transitions, not a universal erasure or security certification.

A later human study would require a separate protocol and appropriate participant governance. It should distinguish perceived usefulness, understanding of disclosure, collaboration behaviour and design outcomes, and account for people nested within teams. A comparison condition and a prospective sample-size rationale would be needed for claims of benefit. No such study, approval, recruitment or outcome is asserted here. For engineering follow-up, the immediate priorities are additional concurrency/failure tests, production-representative query profiling and independent review of the access and data-lifecycle boundaries.

\section{Conclusion}

TeamLens implements voluntary board-scoped self-reporting and descriptive composition within Critsly. Fresh synthetic experiments found exact agreement across the defined 65,536 eligibility subsets and 128 multiplicity fixtures; HTTP and browser checks exercised disclosure, access and withdrawal; and release records connect the implementation to the named Critsly services. The read microbenchmark identified a measurable per-profile membership-query cost. These findings support a bounded account of implementation readiness and reproducibility. Human usefulness, psychometric interpretation, educational effects, comprehensive privacy assurance and production capacity remain separate questions requiring different evidence.

\section*{Research ethics and participant scope}

The reported evaluation used synthetic fixtures, temporary demonstration accounts and software tests. It did not recruit human research participants, administer personality assessments, or collect participant collaboration or learning outcomes. Board-disclosure consent controls are an application feature, not evidence of research-participation consent or institutional ethics approval. No ethics approval, waiver or exemption determination is claimed. Any future study involving participants requires a separate protocol and the applicable institutional review before recruitment or data collection.

\section*{Competing interests}

The author was involved in developing and evaluating Critsly and its TeamLens plugin, the system examined in this report. This involvement is disclosed as a potential competing interest.

\section*{Funding and acknowledgements}

Critsly was developed with support from the AIEDx grant awarded by IAIED (UEN 53487769M).

\section*{Contribution statement}

The contribution comprises the TeamLens integration into the existing Critsly runtime, the disclosure and data-lifecycle design, the executed synthetic evaluation, and the retained engineering evidence. Hendra and colleagues' study motivates the reflection context; its authors did not participate in the present evaluation and are not presented as endorsing TeamLens.

\section*{Data and software availability}\label{sec:availability}

The evaluated source revision is \fullrelease\ in the LearnAdapt deployment repository. A versioned evidence package retained by the author records source/test hashes, commands, environment manifests, synthetic fixture definitions, case results, benchmark trials, analysis outputs and screenshot provenance. Synthetic inputs are identified as such; no real participant-level personality dataset is claimed or deposited. Credentials and session secrets are excluded. The manuscript source package includes an ancillary supplement containing synthetic result records and independent arithmetic-verification scripts. It supports checking the reported aggregation summaries and timing statistics. It does not include the complete production application. At preparation, the full source repository and complete operational evidence package remain in the author's project storage without a public archival identifier. Re-executing the application-level experiments therefore requires access to that pinned runtime; the ancillary checks must not be mistaken for a fresh execution of the service. The live service at \url{https://critsly.com} is not a substitute for the pinned source and retained evaluation files.

\clearpage
\section*{AI assistance}

OpenAI Codex assisted with TeamLens implementation, test-harness development and execution, source inspection, reference verification, analysis, figure preparation and manuscript generation. Unlike a purely documentary adaptation, this work includes newly executed synthetic software experiments and instrumented local timings, identified in the methods and retained evidence. AI assistance did not recruit participants or generate human-outcome observations. No AI system is listed as an author. The named author is responsible for the submitted content, its interpretation and attribution. The use of AI does not change the limited evidential scope of the synthetic tests.

\end{document}